\documentclass[reprint,
 amsmath,amssymb,
 aps,
floatfix,
]{revtex4-1}
\usepackage{amssymb}
\usepackage{dcolumn}% Align table columns on decimal point
\usepackage{amsfonts}
\usepackage{amsmath}
\usepackage{graphicx}
\usepackage{epsfig}
\usepackage{subfigure}
\usepackage[
            pdfstartview=FitH,
            bookmarksnumbered=true,
            bookmarksopen=true,
            colorlinks,
            pdfborder=001,
            linkcolor=blue,
            anchorcolor=green,
            citecolor=red,
            driverfallback=dvipdfm
            ]{hyperref}
\usepackage[toc,page,title,titletoc,header]{appendix}
\usepackage{graphics}
\usepackage{epsfig}
\usepackage{slashed}
\usepackage{latexsym}
\usepackage{syntonly}
\usepackage{amsthm}
\usepackage{epstopdf}
\usepackage{ulem}%ÉŸ³ýÏß
\newcommand{\be}{\begin{equation}}
\newcommand{\bea}{\begin{eqnarray}}
\newcommand{\eea}{\end{eqnarray}}
\newcommand{\ba}{\begin{array}}
\newcommand{\ea}{\end{array}}
\newcommand{\ee}{\end{equation}}
\makeatletter

\newcommand{\Rmnum}[1]{\expandafter\@slowromancap\romannumeral #1@}
\makeatother
\newtheorem{thm}{Theorem}
\newtheorem{mydef}{Definition}
\hfuzz=\maxdimen
\begin{document}
\title{Quantum bit threads of MERA tensor network in large $c$ limit}
\author{Chong-Bin Chen$^{1,2}$}
\thanks{E-mail address: cchongb23@gmail.com}
\author{Fu-Wen Shu$^{1,2}$}
\thanks{E-mail address: shufuwen@ncu.edu.cn}
\author{Meng-He Wu$^{1,2,3}$}
\thanks{E-mail address: menghewu.physik@gmail.com}
\affiliation{
$^{1}$Department of Physics, Nanchang University, Nanchang, 330031, P. R. China\\
$^{2}$Center for Relativistic Astrophysics and High Energy Physics, Nanchang University, Nanchang, 330031, P. R. China\\
$^{3}$Institute of High Energy Physics,
Chinese Academy of Sciences, Beijing 100049, P. R. China}
\begin{abstract}
The Ryu-Takayanagi (RT) formula is a crucial concept in current theory of gauge-gravity duality and emergent phenomena of geometry. Recent reinterpretation of this formula in terms of a set of ``bit threads'' is an interesting effort in understanding holography. In this paper,  we investigate a quantum generalization of the ``bit threads'' based on tensor network, with particular interests in the multi-scale entanglement renormalization ansatz (MERA). We demonstrate that, in the large $c$ limit, isometries of the MERA can be regarded as ``sources'' (or ``sinks'') of the information flow, which extensively modifies the original picture of the bit threads by introducing a new variable $\rho$: density of the isometries. In this modified picture of information flow, the isometries can be viewed as generators of the flow. The strong subadditivity and related properties of the entanglement entropy are also obtained in this new picture. The large $c$ limit implies the classical gravity can be emerged from the information flow.
\end{abstract}

\maketitle
%%\maketitle  IS IGNORED %%%%%%%%%%%
\section{Introduction}
One of the most important developments in AdS/CFT correspondence in the past few years is the discovery of the Ryu-Takayanagi (RT) entanglement entropy formula \cite{Ryu:2006bv}. This formula states that entanglement entropy of a subregion $A$ of a $d+1$ dimensional CFT on the boundary of $d+2$ dimensional AdS is proportional to the area of a certain codimension-two extreme surface in the bulk:
\be
S_A=\frac{\mathrm{area}(m(A))}{4G_N}\label{rt},
\ee
where $m(A)$ is the minimal bulk surface in AdS time slice, which is homologous to $A$, i.e., $m(A)\sim A$. This formula, connecting two important concepts in different fields, suggests some deep relations between quantum gravity and quantum information. Recent progress clearly shows that the RT formula plays a central role in understanding the emergence of space-time.

When exploring the conceptual implications of the RT formula, however, it was firstly noticed by Freedman and Headrick in \cite{bit-threads} that there are some subtleties of the formula. For instance, there is a strangely discontinuous transition of the bulk minimal surface under continuous deformations of $A$. To remove these subtleties, they invoked the notion of ``flow'' which is defined as a divergenceless norm-bounded vector. It turns out that, with the help of the max flow-min cut(MF/MC) principle,  this ``flow'' interpretation of the RT formula is more reasonable: the discontinuous jump disappears and there is more transparent information-theoretic meaning of the properties of the entanglement entropy. In construction of the flow picture of RT formula, the MF/MC theorem plays a crucial role. It roughly states that in some idealized limit, the transport capacity of a classical network is equal to a measure of what needs to be cut to totally sever the network.

The above picture, however, %\sout{is logically incomplete, considering the whole picture}
is built on classical theory of network.
%\sout{More precisely, it seems weird that quantum states (or qubits) are transported by a classical network.}
Recent progress on holography and quantum information theory implies that quantum is playing an important role in the studies of space-time. For example, the TN/AdS correspondence(MERA/AdS, quantum error correction/AdS), the complexity/action correspondence and so on. More reasonable picture should be replaced by a quantum flow network which is tensor network as will see below. In this sense, the above ``flow'' picture of the RT formula is a semi-classical approximation of some unknown quantum (and fundamental) formulations.

For this sake let us move to a tensor network description of quantum physics. Recent study of entanglement in strongly coupled many-body systems has developed a set of real-space renormalization group methods such as the tensor network state representation \cite{Haegeman:2011uy}. In the past few years it has been extensively studied in statistical physics and condensed matter physics. A tensor network description of wavefunctions of a quantum many-body system has a merit to tremendously reduce the number of parameters(from exponential to polynomial) needed in the computation. This makes it a very efficient representation of the wavefunction of the system.  In addition, tensor network representation provides an easy way to visualize the entanglement structure, and the area law of the entanglement entropy is inherent in the network. More attractive property comes from connections between tensor network and the AdS/CFT correspondence \cite{footnote1}, which was first pointed out by Swingle in \cite{Swingle:2009bg}, where he noticed that the renormalization direction along the graph can be viewed as an emergent(discrete) radial dimension of the AdS space. From this perspective, the holography stems from physics at different energy scales and the AdS geometry can be emerged from QFTs \cite{Nozaki:2012zj}. As to the holographic entanglement entropy, the tensor network-based RT formula can be interpreted as sum of all d.o.f of neighbor sites from UV to IR.

Based on these considerations, one natural question is this: what is the ``flow'' picture of the tensor-network based RT formula? In this paper we mainly pay attention to the answer to this question. It turns out that the solution needs the quantum MF/MC(QMF/QMC) theorem which is found recently in \cite{QMF,Cui:2015pla}. This theorem, which is quantum analogy of the MF/MC for tensor networks, states that the quantum max-flow of a tensor network is no bigger than the quantum min-cut of the network. Particularly, for some tensor networks such as MERA \cite{footnote_mera}, the quantum max-flow is equal to the quantum min-cut, in the large central charge $c$ limit.  Based on this theorem and information-theoretic considerations, we define a new variable $\rho$, which can be interpreted as the density of the tensor networks under question. Physically integral of $\rho$ over a region can be viewed as the source (or sink) of the tensor networks and plays a significant role in the flow description of the RT formula. We show that it determines the structure of the tensor network on the basis of a fixed Lorentzian manifold $(\mathcal M,g)$. More precisely, $dV_{network}=\rho (x)\sqrt{g}dV$ of this tensor network.
When $\rho(x)=$constant, it reduces to the kinematic space of an $AdS_3$ time-slice \cite{Czech:2015kbp,Czech:2015qta} and the corresponding tensor network is MERA.
%\sout{In addition, from the information viewpoint a tensor network is a quantum circuit that maps a reference state to a target state by the network and quantum gates(tensors). In this language, $\rho$ is the density of compression or decompression of quantum bits through reducing or expanding the dimensions of Hilbert space. Specifically, MERA tensor network where $\rho=$constant has $dS_2$ geometry, if we regard MERA as kinematic space of an $AdS_3$ time-slice as first pointed out in \cite{Czech:2015kbp}. In this case, our formulation suggests a naive picture that the evolution of our space-time can be regarded as a huge and complex quantum circuits and expanding is a progress that decompresses and entangles quantum bits continuously.}
In addition, from the informaton viewpoint $\rho(x=(i,j))$ is the density of compression or decompression of quantum bits through reducing or expanding the dimensions of Hilbert space. It encodes how much information shares between two sites $i$ and $j$. In other word, $\rho(x)$ provides the local contributions of these sites to the total conditional mutual information.

The organization of this paper is the following. In section 2 we first give a brief review of Freedman-Headrick's proposal of bit threads and holography, followed by a brief introduction of the QMF/QMC theorem. To find out the relation between the QMF/QMC theorem and the RT formula for MERA tensor network, in this section we also have reviewed the MERA on kinematic space. In section 3 we briefly import our main results from informational points of view. In section 4 we propose our ``flow'' language of a tensor network with the help of QMF/QMC theorem. In section 5 we give an information-theoretic interpretation of the MERA, with particular interests in the information-theoretic meaning of the isometry. Several physical key points of the picture are also discussed in this section. In last section we draw our main conclusions and discussions.

\section{Background Setup}\label{0002}

\subsection{Bit threads and holography}
\begin{figure}
\centering
\includegraphics[scale=0.5]{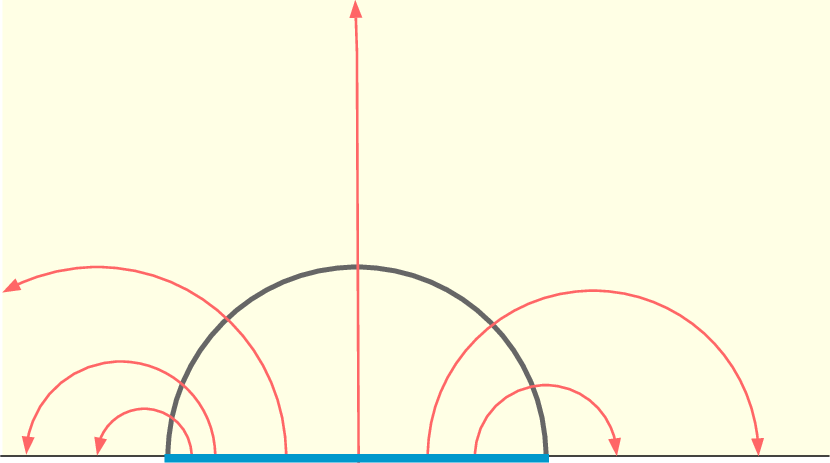}
\caption{(Color online) Max-flow Min-cut picture of a subregion. Red line is the bit threads which have a maximal density on a minimal surface(black curve). }\label{bit-threads}
\end{figure}

The RT formula \eqref{rt} can be reinterpreted as a ``max-flow'' from max-flow/min-cut(MF/MC) theorem in Riemannian geometry firstly explored in \cite{bit-threads}. And can be understood as calibrations to Ryu-Takayanagi minimal surfaces \cite{Bakhmatov:2017ihw}. To understand this point, we define a divergenceless vector field $v$ as a ``flow'' satisfying the following two properties \cite{bit-threads} (see FIG.\ref{bit-threads}):
\begin{eqnarray}
|v| \le C,\\
\nabla_\mu v^\mu = 0,
\end{eqnarray}
where $C$ is a positive constant. Then the flux of $v$ through an oriented manifold surface $m \sim A$ can be defined as an integral:
\begin{eqnarray}
\int_{m(A)} v := \int_{m(A)} \sqrt {h}n_\mu v^\mu,
\end{eqnarray}
where $h$ is the determinant of the induced metric on $m$ and $n_\mu$ is the unit normal vector. The maximal flux should be bounded by a bottleneck
\begin{eqnarray}\
\int_A v = \int_{m(A)} v\le C\int_{m(A)} \sqrt {h} = C\ \mathrm{area}(m).
\end{eqnarray}
This inequality is saturated by C where $n_\mu v^\mu = C$ holds. This indicates that a flow reaches its maximum if and only if $n_\mu v^\mu = C$, and the maximal flux is equal to the minimal area multiplying a constant:
\begin{eqnarray}
\max_v \int_A v = C\ \min_{m\sim A} \mathrm {area}(m).
\end{eqnarray}

It is necessary to introduce two extensions of the theorem. Firstly, when we change $A$ continuously the maximal flow $v(A)$ also varies continuously. Secondly, consider two disjoint regions $A$ and $B$ of the boundary, in general we cannot find a flow which maximizes the flux through $A$ and $B$ simultaneously, i.e.
\begin{eqnarray}
\int_A v + \int_B v &=& \int_{AB} v \le C\mathrm{area}(m(AB)) \nonumber \\
                    &<& C\ \mathrm{area}(m(A)) + C\ \mathrm{area}(m(B)).
\end{eqnarray}
We call this  ``nesting'' property.

Now return to holography. One can replace the minimal area with the maximal flow and the RT formula can be rewritten in the following way:
\begin{eqnarray}
S(A) = \max_v \int_A v,
\end{eqnarray}
where $C = 1/(4G_N)$. Recall that a magnetic field is visualized as field lines in common. Similarly, these flow lines $v$ can be regarded as oriented ``bit threads'' from boundary to bulk. The upper bound $1/(4G_N)$ of flow can be interpreted as this: the bit threads cannot be tighter than one per $4$ Planck areas in $1/N$ effects. Then a thread which emanates from boundary region $A$ should be viewed as one independent bit of information carrying out of $A$. From this point of view the maximal number of independent information is the entanglement entropy $S(A)$.

From this ``flow'' language, we can also obtain the conditional entropy and mutual information. Let $v(A;B)$ denote the flow which not only maximizes the flux through $A$ but also maximizes the flux through $AB$, i.e. $v(A)$ or $v(A,B)$. Then the conditional entropy $H(A|B):=S(AB)-S(B)$ can be rewritten as an expression in terms of flows \cite{bit-threads}. Consider two regions case, the conditional entropy can be written as
\begin{eqnarray}
H(A|B) &=& \int_{AB} v(B;A) - \int_B v(B;A) \\
       &=& \int_A v(B;A).
\end{eqnarray}
At the same time, flux through $A$ reaches its minima. From this, one can also write down the mutual information $I(A:C) := S(A)-H(A|C)$. Without loss of generality, one can choose the entropy of $A$: $\int_A v(A;C)$, then $I(A:C)$ can be rewritten as
\begin{eqnarray}
I(A:B) &=& \int_A v(A;B) - \int_A v(B;A) \\
       &=& \int_A (v(A;B) - v(B;A)).
\end{eqnarray}
This is the flux which can be shifted between $A$ and $C$. Similarly, in three regions case, we can also write down the conditional mutual information $I(A:C|B) := S(AB)+S(BC)-S(ABC)-S(B) = H(A|C)-H(A|BC)$. Without loss of generality, one can choose $H(A|B) = \int_A v(B,A;C)$ and $H(A|BC) = \int_A v(B,C;A)$, in this way $I(A:C|B)$ can be expressed as:
\begin{eqnarray}
I(A:C|B) &=& \int_A v(B,A;C) - \int_A v(B,C;A) \\
         &=& \int_A (v(B,A;C) - v(B,C;A)).
\end{eqnarray}

\subsection{Quantum Max-flow/Min-cut} \label{section_2.2}
 The quantum max-flow min-cut(QMF/QMC) conjecture was first presented in \cite{QMF}. Then in \cite{Cui:2015pla} Cui et al. showed that this conjecture dose not hold in general, except for some given conditions. There are two versions of this conjecture, we first review the first version.

\begin{figure}
\centering
\includegraphics[scale=0.4]{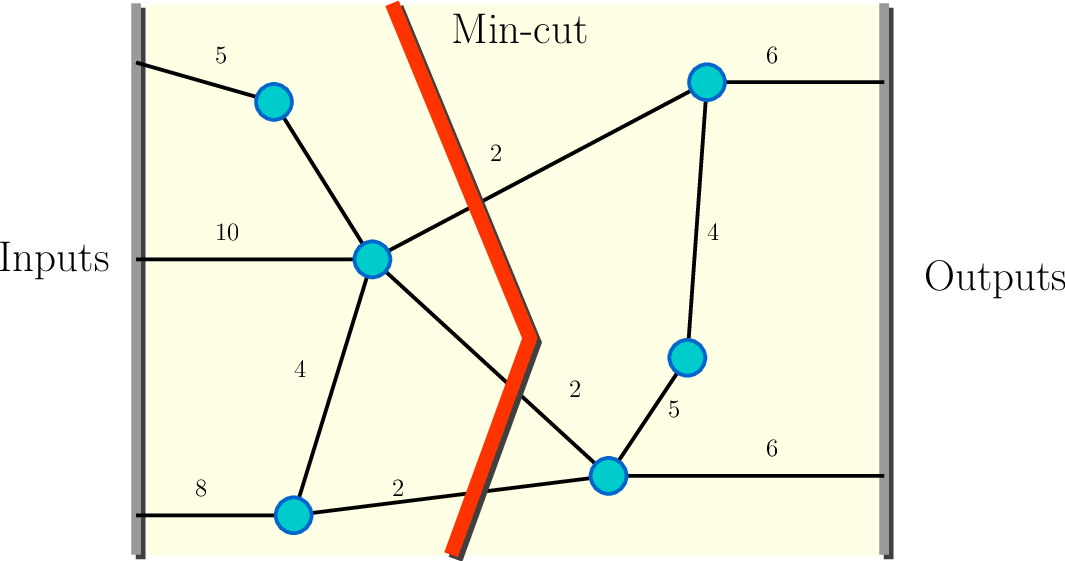}
\caption{(Color online) A tensor network which has three inputs and two outputs. The blue balls are vertexes associating tensors and the black lines are edges associating different Hilbert dimensions. The red line is the minimal cut of this network. }\label{mfmc}
\end{figure}

Tensor network can be regarded as a graph $G(\tilde V,E)$ which is unoriented, meanwhile has a set of inputs $S$ and a set of outputs $T$. $\tilde V$ is a disjoint partition $\tilde V = S \bigcup T \bigcup V$. $E$ is a set of edges with a capacity function $a: E\to \mathbb N, e \mapsto a_e$. In tensor network, each edge $e$ is associated with a Hilbert space $\mathbb C^{a_e}$, and the capacity of edges is dimensions of the corresponding Hilbert space. Inputs $S$ and outputs $T$ can be thought as some open edges with open ends for convenience. $V$ is a set of vertices that for each vertex $v$ there are $d_v$ edges $e(v,1), e(v,2), \cdots , e(v,d_v)$ incident to $v$. Associating a tensor to every vertex $v\mapsto \mathcal T_v \in \mathcal I^v:=\bigotimes_{i=1}^{d_V} \mathbb C^{a_e}$ thus sends a graph $G$ to a tensor network,$G \mapsto N(G,a;\mathcal T)$ (see an example in FIG.\ref{mfmc}). After fixing basis of the Hilbert space of the open edges, we can determine a state $|\alpha(G,a;\mathcal T)\rangle \in V_S \bigotimes V_T$ which is given by
\begin{eqnarray}
|\alpha(G,a;\mathcal T)\rangle := \sum_{\substack{i_1,\cdots,i_{|S|}\\j_1,\cdots,j_{|T|}}} &C&_{i_1,\cdots,i_{|S|},j_1,\cdots,j_{|T|}} \\
&\times&|i_1,\cdots,i_{|S|}\rangle_S  |j_1,\cdots,j_{|T|}\rangle_T,\nonumber \label{2.2.1}
\end{eqnarray}
where $V_S:=\bigotimes_{u\in S}\mathbb C^{a_{e(u)}}$ and $V_T:=\bigotimes_{u\in T}\mathbb C^{a_{e(u)}}$. With these preparations, one can start to define a quantum max flow and min cut.

Before that, the definition of ``cut'' must be stated. If there exists a partition $\tilde V = \bar S \bigcup \bar T$ so that $S\subset\bar S, T\subset\bar T$, then a cut $A$ is a set that $A = \{(u,v) \subset E:u\in \bar S,v\in \bar T\}$. Intuitively, removing the edges in $C$ will lead to disconnect path from $S$ to $T$.
\begin{mydef}[Quantum Min-cut]
The quantum min-cut $QMC(G,a)$  is the minimum value of product of capacities over all edge cut sets, i.e.
\begin{eqnarray}
QMC(G,a) := \min_A\ \prod_{e \in C} a_e.
\end{eqnarray}
\end{mydef}

There is a linear map $\beta(G,a;\mathcal T) \in V_S^* \bigotimes V_T = Hom(V_S, V_T)$ from inputs to outputs: $V_S \mapsto V_T$ acting on the inputs state:
\begin{eqnarray}
\beta(G,a;\mathcal T)|i_1,\cdots,i_{|S|}\rangle_S :&=& \sum_{j_1,\cdots,j_{|T|}} C_{i_1,\cdots,i_{|S|},j_1,\cdots,j_{|T|}}\nonumber \\
&\times&|j_1,\cdots,j_{|T|}\rangle_T.
\end{eqnarray}
It is obvious that in this basis the matrix $C$ is exactly the $\beta (G,a;\mathcal T)$. Then one can define the quantum max-flow as follows:
\begin{mydef}[Quantum Max-flow]
For over all tensor assignments, there exists a maximal value of the rank of map $\beta(G,a;\mathcal T)$ and we define this maximal value as the quantum max-flow:
\begin{eqnarray}
QMF(G,a) := \max_{\mathcal T}\ rank(\beta(G,a;\mathcal T)).
\end{eqnarray}
\end{mydef}

Cui et al. \cite{Cui:2015pla} stated that $QMF(G,a)$ is not equal to $QMC(G,a)$ in general. In fact the $QMF(G,a)$ is always no larger than $QMC(G,a)$ in a given finite graph $G$: $QMF(G,a)\le QMC(G,a)$. The equality holds in a special case that can be considered as a weak QMF/QMC:
\begin{thm}[Quantum Max-flow Min-cut theorem]
For a given graph $G(\tilde V,E)$, if the capacity $a$ of each edge is a power of $d$, where $d>0$ is an integer, then
\begin{eqnarray}
QMF(G,a) = QMC(G,a).\label{QMF=QMC}
\end{eqnarray}
\end{thm}

Now let us turn to the entanglement entropy between inputs and outputs of tensor network and see its relation with the QMF and QMC. The Hilbert space of pure sates (\ref{2.2.1}) is $\mathcal H = V_S \bigotimes V_T$ and one can obtain the reduced density matrix of $|\alpha(G,a;\mathcal T)\rangle$ on $S$ by tracing $T$:
\begin{eqnarray}
\rho_S\left(\frac{|\alpha(G,a;\mathcal T)\rangle}{\sqrt {Tr(CC^\dag)}}\right) &=& Tr_T\ |\alpha(G,a;\mathcal T)\rangle \langle \alpha(G,a;\mathcal T)|\nonumber \\
&=& \frac{CC^\dag}{Tr(CC^\dag)},
\end{eqnarray}
where $Tr(CC^\dag) = \sum_{i_1,\cdots,i_{|S|},j_1,\cdots,j_{|T|}}|C_{i_1,\cdots,i_{|S|},j_1,\cdots,j_{|T|}}|^2$, and $\frac{|\alpha(G,a;\mathcal T)\rangle}{\sqrt{Tr\ CC^\dag}}$ is a normalized state. We have already known the von Neumann entropy is $S(\rho):=-Tr\ (\rho\log \rho)$. Now define an entanglement entropy between $S$ and $T$:
\begin{align}
\begin{split}
EE(G,a;\mathcal T) &:= S\left(\frac{CC^\dag}{Tr(CC^\dag)}\right)\\
                   & = -\frac{Tr(CC^\dag\log(CC^\dag))}{Tr(CC^\dag)}+\log(Tr(CC^\dag)).\label{EE}
\end{split}
\end{align}
 Let $MEE(G,a)$ denote the maximal of $EE(G,a)$ over all $\mathcal T's$. One can prove that in general $MEE(G,a) \le \log\ QMF(G,a) \le \log\ QMC(G,a)$.
The equality holds when one considers the same case as theorem 1, i.e.
\begin{thm}
For a given graph $G(\tilde V,E)$, if the capacity $a$ of each edge is a power of $d$, where $d>0$ is an integer, then
\begin{equation}
MEE(G,a) = \log\ QMC(G,a) = \log\ QMF(G,a).\label{EE2}
\end{equation}
\end{thm}

The second version of QMF/QMC conjecture is more restricted. The vertices of the same type have to be assigned the same tensor. More specifically, we put an ordering $O$ to the ends of the edges incident to each vertex and define a $valence\ type\ B_v$ of a vertex $v$ to be the sequence $(a_{e(v,1)},..., a_{e(v,d_v)})$, where $a_{e(v,d_v)}$ is dimensions of the Hilbert space of edge $e(v,d_v)$. Let us denote $\mathcal B(G,a,O)$ as the set of valence type of vertices of graph $G$. Now the vertices with the same valence type have to be assigned the same tensor $\mathcal T = \{\mathcal T_B:B\in\mathcal B(G,a,O)\}$.

From above we also obtain a linear map, which denoted by $\beta(G,a,O;\mathcal T)$ in $Hom(V_S,V_T)$. The definition of quantum max-flow $QMF(G,a,O)$ for this second version is the maximum rank of $\beta(G,a,O;\mathcal T)$, similar to the first version.
The difference is here: Conditions stated in Theorem 1 are insufficient to guarantee the equality $QMF=QMC$, even when the Hilbert space dimension is same in each edge in the graph. The restriction of tensor, or the quantum gate of circuit is more nontrivial than the first version of QMF/QMC. We have known the tensors of MERA network are restricted and the second version of QMF/QMC is more suitable for our discussion. We still need some additional conditions and we will discuss it latter.

\subsection{MERA on kinematic space}
\begin{figure}
\centering
\includegraphics [scale=0.4]{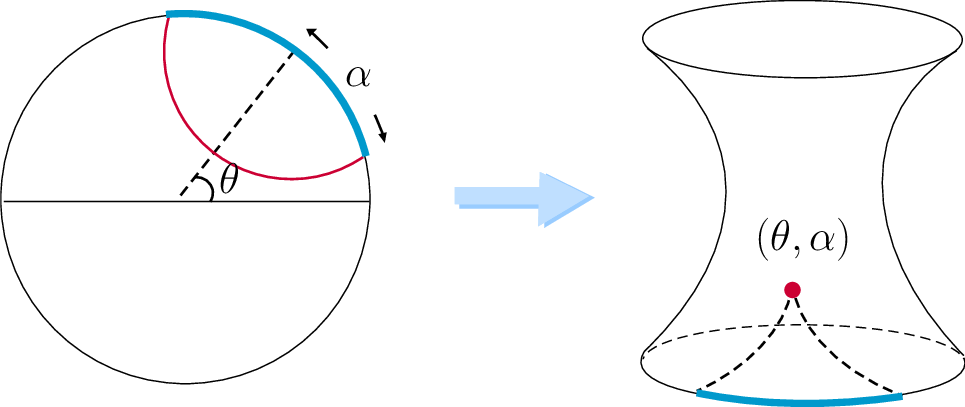}
\caption{A space-like geodesic on AdS$_3$ timeslice can be mapped to a point $(\theta,\alpha)$ in kinematic space.}\label{kin}
\end{figure}
For the following convenience let us give a brief review on kinematic space which was firstly formulated in \cite{Czech:2015qta,Czech:2015kbp}. Given a hyperbolic plane $\mathbb H^2$ which is a time slice of pure $AdS_3$ space-time:
\begin{eqnarray}
ds^2 = d\rho^2 + \sinh^2 \rho d\tilde \theta^2.
\end{eqnarray}
The equation of a space-like geodesic that anchors on boundary points is:
\begin{eqnarray}
\tanh \rho \cos(\tilde\theta - \theta) = \cos \alpha,
\end{eqnarray}
where $(\theta,\alpha)$ are parameters which label a oriented geodesic as shown in FIG.\ref{kin}. Space of all these geodesics $(\theta,\alpha)$ forms a 2-dimensional manifold which is called $kinematic\ space$.  A geodesic can be described by a point $(\theta,\alpha)$ in the kinematic space. Crofton's formula in $\mathbb H^2$ states that the length of a curve $\gamma$ can be measured by the number of the geodesics which intersect it, i.e.
\begin{eqnarray}
length\ of\ \gamma = \frac{1}{4}\int_Kn(g,\gamma)\mathcal Dg,
\end{eqnarray}
where $n(g,\gamma)$ is the number of geodesics intersect $\gamma$ and $\mathcal D g$ is the measure on the kinematic space
\begin{eqnarray}
\mathcal Dg = \frac{\partial^2 S(u,v)}{\partial u \partial v}dudv. \label{2.3.4}
\end{eqnarray}
If the curve is a geodesic with two ends $u$ and $v$ anchoring on boundary then $S(u,v)$ is the length of the geodesic . We have used a light-cone coordinate on the kinematic space,
\begin{eqnarray}
u = \theta - \alpha\ \ \ \ \  and\ \ \ \ \  v = \theta + \alpha.
\end{eqnarray}
Eq. (\ref{2.3.4}) is also the line element of the kinematic space multiplied by some coefficient
\begin{eqnarray}
ds^2_{kinematic} \sim \frac{\partial^2 S_{ent}(u,v)}{\partial u \partial v}dudv. \label{2.3.6}
\end{eqnarray}
where we have replaced $S$ by $S_{ent}$ (the entanglement entropy) because of the RT formula. Therefore the entanglement entropy can be represented by a volume in kinematic space.
\begin{eqnarray}
S_{ent} &=& \frac{length\ of\ \gamma}{4G}\nonumber \\
        &=& \frac{1}{4}\int_K n(g,\gamma) \frac{\partial^2 S_{ent}(u,v)}{\partial u \partial v}dudv.
\end{eqnarray}

Czech et al.\cite{Czech:2015kbp} argued that, considering the auxiliary causal structure of MERA, it is the vacuum kinematic space instead of the $AdS_3$ time slice that should be viewed as the corresponding geometry of the MERA. So the kinematic space becomes an intermediary in the AdS/CFT.
\begin{figure}
\centering
\includegraphics[scale=0.55]{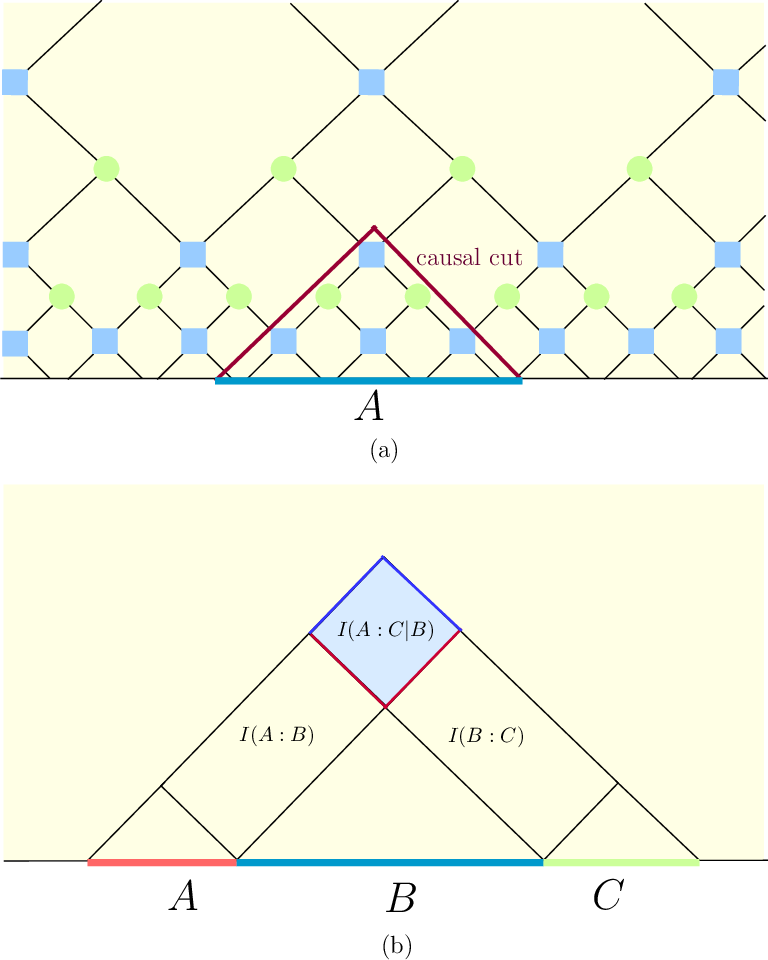}
\caption{(a) A causal cut of MERA tensor network. (b) Conditional mutual information $I(A:C|B)$ on boundary can be regarded as volume of a causal domain in kinematic space.}\label{mera}
\end{figure}

One of the key points of this argument is the casual structure of the MERA. It turns out that this makes it more natural to match such a network with a Lorentzian manifold, as first mentioned in \cite{Beny:2011vh}. To proceed, let us consider exclusive causal cone for part of lattices. Boundary of this causal cone is called as causal cut of the MERA network as shown by the red line in FIG.\ref{mera}(a). The method to calculate the entanglement entropy -- given a holographic interest in MERA in \cite{Swingle:2009bg} -- is just counting the number of edges cut by the causal cut in this network, with each edge assigning a weight $\log\chi$, where $\chi$ is the Hilbert dimension of edges. In the same footing, one can also calculate the conditional mutual information by counting the number of edges. Recall that conditional mutual information is defined as follows:
\begin{eqnarray}
I(A:C|B) = S(AB) + S(BC) - S(ABC) - S(B).
\end{eqnarray}
It means that the conditional mutual information can be obtained by counting the number of edges which is the net reduction of edges through a causal diamond  from bottom up as shown in FIG.\ref{mera}(b). One may find that this is the same as counting the number of isometries living in this diamond, because every isometry has two input edges and only one output edge from bottom up. In other words, each isometry soaks up an edge so that counting the number of it is precisely equal to counting the decrease in the number of edges. It shows that the conditional mutual information is proportional to the number of the isometries in causal diamond, and is also proportional to the volume of this diamond which can be easily seen from (\ref{2.3.6}).

Based on these observations, connections between conditional mutual information and volume in kinematic space can be built. Czech et al.\cite{Czech:2015kbp}  adopted conditional mutual information as a definition of volume in MERA,
\begin{eqnarray}
\mathcal{D}(isometries)=I(A:C|B).
\end{eqnarray}
This formula evaluates the amount of the volume after compressing the state living on its past edges. In this vacuum MERA, the `density of compression' is proportional to the number of isometry. More explicitly, the metric of a discrete tensor network is given by
\begin{eqnarray}
ds_{network} &= I(\Delta u,\Delta v|B)\\
             &\xrightarrow{MERA}& (\#\ of\ isometries)\Delta u\Delta v \nonumber,
\end{eqnarray}
and this is the metric of kinematic space, also the volume element in kinematic space.

\section{A information-theoretic viewpoint}\label{iii}
\begin{figure}
\centering
\includegraphics [scale=0.55]{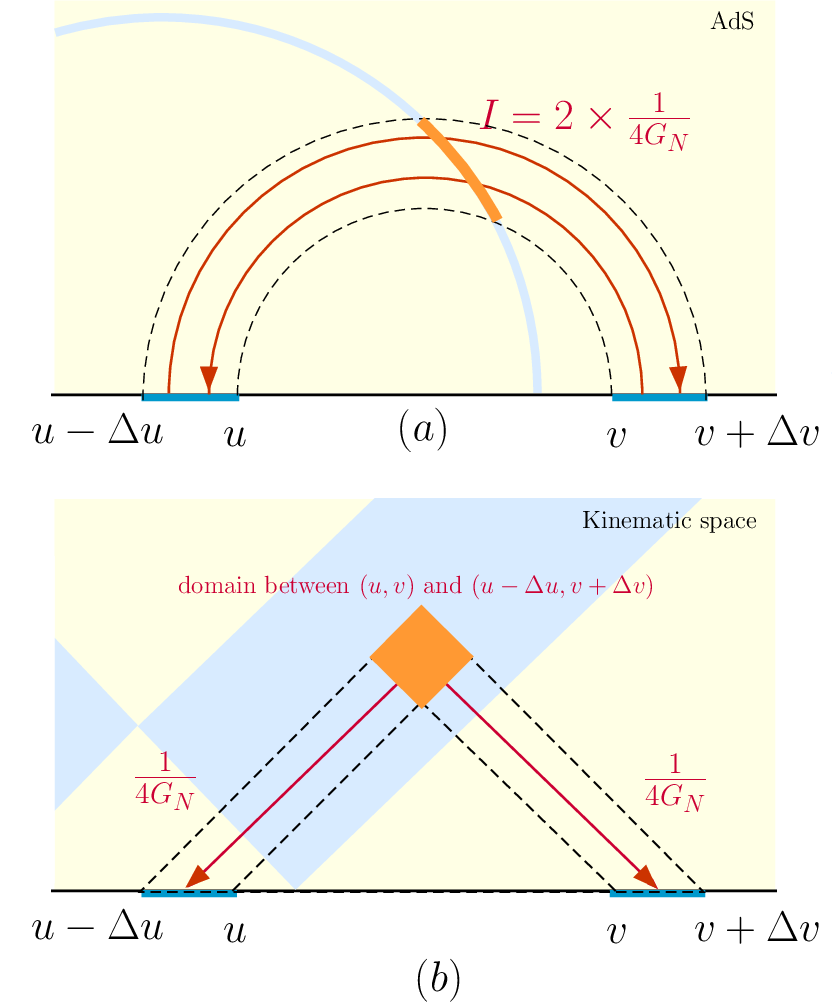}
\caption{The conditional mutual information between $[u-du,u]$ and $[v,v+dv]$ which is located in minimal surface can be interpreted as $2\times 1/4G_N$ decompressed information in kinematic space.}\label{information}
\end{figure}

As stated in the previous section, we prefer to treat MERA as a discrete kinematic space rather than the original slice of AdS space. This statement is based on the following two advantages: $(a)$ they share the same causal structure and $(b)$ regarding entanglement as ``flux'' through causal cut, which is equal to count the number of lines on causal cut, has more natural interpretation in kinematic space
%\sout{\textcolor{blue}{.The counting of entanglement flow is equal to counting the number of lines on causal cut} }
\cite{Czech:2015kbp}. As a consequence, bit threads in AdS time slice should have an information-theoretic interpretation on kinematic space. To see this, we first recall flows in AdS time slice. A flow $v$ is a vector field and one can define a set of integral curves of $v$ whose transverse density equals $|v|$. Each flow line, the so-called ``bit threads'', is an oriented thread connecting two different points on the boundary. For example, given a time slice of AdS, we can split boundary into two parts $A$ and $A^c$, then the information (flow) shared between $A$ and $A^c$ is
\begin{equation}
I(A:A^c) = 2S(A).\label{IS}
\end{equation}
A thread between $A$ and $A^c$ connects two points on $A$ and $A^c$, of which one is the start point of thread (belongs to $A$) and the other is the end point (belongs to $A^c$). These two end points can be mapped to a point in the kinematic space, which is denoted by $(u, v)$. We therefore have a picture between original space and kinematic space, as sketched in FIG.\ref{information}.

In the previous section we have mentioned that one of important properties of the bit threads is $|v|\le 1/(4G_N)$, which means that one cannot contain more than $ 1/(4G_N)$ information in unit area. The flow density is saturated at the minimal surface $m(A)$ , i.e, $|v| = 1/(4G_N)$. This indicates that we have
%\sout{$1/4G_N$ bits information per unit area on the surface.}
$1/(4G_N)$ entanglement flow per unit area on the surface. From (\ref{IS}), we can think $A$ and $A^c$ share $2 \times 1/(4G_N)$ bits of information per unit areas of minimal surface, or equivalently, $[u-\Delta u, u]$ and $[v, v+\Delta v]$
%\sout{shared $2 \times 1/(4G_N)$.}
on the boundary share $2\times1/(4G_N)$ bits of information. Mapping this ``area'' (actually is a geodesic length in 2d) to kinematic space yields a ``volume'' ( an area in 2d) of some region as shown in FIG \ref{information}. This implies that $2 \times 1/(4G_N)$ bits of information in unit area of minimal surface correspond to $2 \times 1/(4G_N)$ bits of information in $\Delta u\Delta v$ in kinematic space(we set the radius of kinematic space to 1).
%\sout{, which is equivalent to count the number of isometries in this domain.}
Regarding to these partial contributions to the total $I(A:A^c)$, the relation  between AdS and kinematic space is given by the following graph:
\begin{figure}[!htbp]
\centering
\includegraphics [scale=0.7]{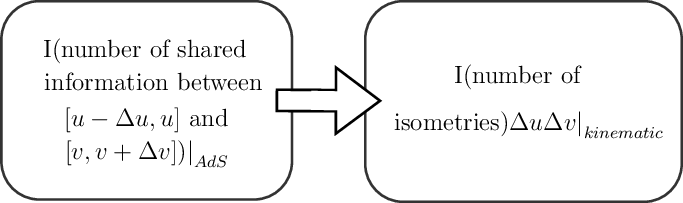}
\end{figure}

This point can be also seen from the causal structure perspective. The shared information between $[u-\Delta u,u]$ and $[v, v+\Delta v]$  can be naively regarded as some non-intersect geodesics included in the tube $([u-\Delta u,u], [v, v+\Delta v])$ in AdS \cite{Agon:2018lwq}, which are time-like between $[u,v]$ and $[u-\Delta u, v+\Delta v]$. In kinematic space, these are  isometries in a causal diamond between $(u,v)$ and $(u-\Delta u, v+\Delta v)$, which, as expected, is the region where the conditional mutual information $I([u,u-\Delta u]: [v,v+\Delta v]|[u,v])$ is calculated. $I([u,u-\Delta u]: [v,v+\Delta v]|[u,v])$, as a part of the total $I(A:A^c)$, is only related to the degrees of freedom between $[u-\Delta u,u]$ and $[v,v+\Delta v]$ . Therefore we should introduce a local quantity associated with the isometries of the MERA in accordance with the above discussions.

Above analysis strongly suggests to regard the isometries as ``sources'' (or ``sinks'') of information. Indeed, if we deem the information flow $1/4G_N$ from up to down in MERA through isometries, then these isometries decompress
%\sout{it. The decompression results in $2\times 1/(4G_N)$ bits information}
these bits. The decompression result is $2\times 1/(4G_N)$, which is the conditional mutual information between $[u-\Delta u, u]$ and $[v, v+\Delta v]$. And the isometries play roles in sharing the information in these two intervals on the boundary as shown in FIG.\ref{information}. That is to say, isometries provide a local ``density of compression(or decompression)'' of network and such MERA network can be regarded as an iterative compression algorithm which maps the density matrix of a interval to a compressed state on causal cut \cite{footnote:entangler}.

In this paper we try to further study the holography of tensor network from a different perspective: the QMF/QMC theorem. We discuss the quantum version of network which is different from the classical one of Headrick-Freedman. We try to set up a picture of holography of quantum network and study some general properties between tensor network and space-time. The main results of this work are following:
\begin{itemize}
\item [1)]
For a quantum network, we introduce a density $\rho(x)$ to describe the QMF of a general tensor network. We focus on the MERA which has some important properties (such as symmetry) for holography. We find from QMF=QMC case that each edge has the same entanglement flow (assume all edge have the same Hilbert dimension) and hence if there exists isometry in the tensor network, the source (or sink) has to be introduced to preserve conservation of entanglement flow. Therefore, the quantum bit threads is studied in the kinematic space and we extend the picture of B. Czech et al. by introducing such $\rho$.
\item [2)]
We also realize that in MERA, the density $\rho$ encodes the local contributions of each degree of freedom to the total conditional mutual information between two intervals on the boundary. This is similar to the so-called entanglement contour studied in holography recently \cite{Kudler-Flam:2019oru,Vidal:contour}. We also find the definition of $\rho$ meets some requirements similarly to the contour.
\item [2)]
We argue in MERA tensor network, which has the same type isometry everywhere, the ``classical'' limit QMF=QMC is equivalent to the classical limit of emerged space-time. This is the large central charge limit of tensor network dual to classical gravity, which lacks of discussion in other articles.
\end{itemize}

\section{Towards a Tensor Network/Flow correspondence}
\subsection{QMF=QMC implies a isometric tensor}\label{isometric_tensor}
In this subsection we consider the QMF$=$QMC case. Under this condition, the tensor between a causal cut and boundary is an isometric tensor which is defined in \cite{Pastawski:2015qua}. To see this, we set the causal cut as the inputs $S$ of network and the corresponding boundary region as the outputs $T$ of network. Now identifying $V_S^*$ with $V_S$ by using the chosen basis in $V_S$, one can determine a state $\alpha(G,a;\mathcal{T})$. We denote the basis of $V_S, V_T$ as ${|i\rangle_S},{|j\rangle_T}$ respectively and let matrix of $\beta(G,a;\mathcal{T})$ be $C$ under these basis, we have
\begin{equation}
\alpha\left(G,a;\mathcal{T} \right)=\sum_{i,j}C_{ji}|j\rangle_T|i\rangle_S.
\end{equation}
In other words, $\beta$ is a tensor which maps from causal cut to boundary:
\begin{equation}
\beta:~~|i\rangle_S~\mapsto~\sum_{j} C_{ji}|j\rangle_T
\end{equation}
We assume that $S$ is a minimal cut and $\dim(S)\le\dim(T)$. If QMF/QMC conjecture holds in this tensor network QMF$=$QMC. Then after an appropriate ordering of the basis elements in $V_S$ and $V_T$, the map $\beta$ has such simple form \cite{Cui:2015pla}:
\begin{equation}
\left(
\begin{array}{ccc|c}
\begin{matrix}
1&~~&~~\\
~~&\ddots&~~\\
~~&~~&1\\ \hline
~~&0&~~\\
\end{matrix}
\end{array}\right)
\end{equation}
Let $M=\text{QMC}(G,a)$ be the dimension of inputs, the upper block matrix is $M\times M$. So we have
\begin{equation}
\sum_{j}\beta^{\dag}_{i'j}\beta_{ji} = \delta_{i'i}.
\end{equation}
That is, $\beta$ is the so-called isometric tensor. The difference is that the tensor talked in \cite{Pastawski:2015qua} is defined in a negative curvature space but ours is in kinematic space, which is a positive curvature space. Such tensor network has an important property: the RT formula holds $S_{EE}=|S|\cdot\log \chi$. It implies
%\sout{the bipartition of network have maximal entanglement between two parts of this network}
after a bipartition of network, two parts of this network have maximal entanglement. ($|S|$ is the number of cut legs and $\chi$ is the Hilbert dimension of each edge). From (\ref{EE}) we have known the entanglement entropy is given by the tensor $C$(or $\beta$), which is more related to QMF. If the QMF/QMC asymptotically holds, the second term in (\ref{EE}) will be leading and the entropy is given by the QMC (\ref{EE2}). This means in this case, each edge through bipartition cut has maximal entropy flow $\log \chi$. In the next subsection our model is considered in the QMF$=$QMC case and more interpretation about this will be showed in section \ref{5.A}.

\subsection{General setup}
We start by explaining a tensor network in terms of ``flow'' language that is convenient for our discussion. Suppose there is a flow through a boundary, which can be denoted by $f^\mu$. Its flux $\mathcal F$ through a boundary region $A$ is obtained by integration $f^\mu$ over this region:
\begin{eqnarray}
\mathcal F = \int_{A} f \ : =  \int_{A} \sqrt {|h|}n_\mu f^\mu. \label{a.1}
\end{eqnarray}
where $h$ is the determinant of the induced metric on $A$(Actually because of the IR divergence we should choose an IR cut-off surface $A_{\epsilon}$, but for simplification we still use $A$ to denote it). It is obvious that it satisfies the additivity
\begin{eqnarray}
\int_{A} f  + \int_{B} f = \int_{AB} f.
\end{eqnarray}

Now let us consider a cut $C_A\sim A$ to be an oriented codimension-one submanifold in network which is homologous to A. As claimed in the last section, isometries in the tensor network may play a role of source (or sink). Therefore, different from bit threads in original AdS slice, in general $\int_{A} f \neq \int_{C_A} f $. Instead, one extra term which describes the contributions from tensors should be added
\begin{eqnarray}
\int_{C_A} f = \int_{A} f + \int_{D_{A}} \rho. \label{3.1.2}
\end{eqnarray}
where $\int_{D_A} \rho := \int_{D_A}\rho\sqrt {-g}d \omega$ and $g$ is the determinant of the metric on this Lorentzian manifold. $D_A$ is the region enclosed in $A\bigcup C_A$.
%\sout{$\rho$ can be viewed as density of tensors and  should satisfy the following two properties:}
%\begin{eqnarray}
%\sout{\ |\rho| \le \rho_M,}\\
%\sout{\nabla_\mu f^\mu=-\rho,} \label{divergencelessnes}
%\end{eqnarray}
%\sout{where $\rho_M$ is a positive constant. The first constraint implies finiteness of the density.}
From \eqref{3.1.2}, it is straightforward to get
\begin{eqnarray}
0&=& -\int_{AB} f - \int_{BC} f + \int_{ABC} f + \int_{B} f \nonumber \\
&=& \int_{D_{AB}} \rho + \int_{D_{BC}} \rho - \int_{D_{ABC}} \rho - \int_{D_{B}} \rho \nonumber \\
&+& \int_{C_{AB}} f + \int_{C_{BC}} f - \int_{C_{ABC}} f - \int_{C_B} f \nonumber \\
&=& \int_{D} \rho + \oint_{\partial D} f. \label{a.5}
\end{eqnarray}
where $D$ is the region $D_{ABC}+D_B-D_{AB}-D_{BC}$. That means for an arbitrary region in a network the Gauss' theorem is always tenable.
$\rho$ can be viewed as density of tensors and  should satisfy the following two properties:
\begin{eqnarray}
\ |\rho| \le \rho_M,\\
\nabla_\mu f^\mu=-\rho, \label{divergencelessnes}
\end{eqnarray}
where $\rho_M$ is a positive constant. The first constraint implies finiteness of the density.
The second term of last equality in (\ref{a.5}) is the flux and can be denoted by $\mathcal D$. Eq.  (\ref{a.5}) indicates that the flux $\mathcal{D}$ of the region $D$ can be also calculated by a volume integral instead of a surface integral
\begin{eqnarray}
\mathcal{D} :=\oint_{\partial D} f= -\int_{D} \rho . \label{3.1.6}
\end{eqnarray}
This implies that a flow is incoming from the bottom of the casual diamond. Meanwhile, the constraint $|\rho| \le \rho_M$ implies the flux is bounded by
\begin{eqnarray}
\left| \mathcal{D} \right| \le \rho_M\int_{D} \sqrt {|g|}d \omega = \rho_M V_D. \label{a.7}
\end{eqnarray}

\begin{figure}
\centering
\includegraphics [scale=0.55]{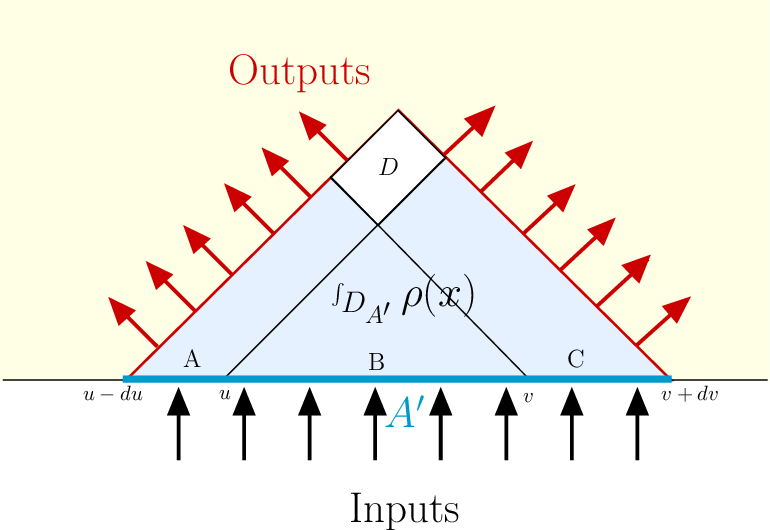}
\caption{We set the cutoff legs as inputs $A'$ and the causal cut legs as outputs $C_{A'}$. The blue triangle(including region $D$) is $D_{A'}$. For a general network, one can find out the causal cut of intervals $A,B,C,AB,BC$ and $ABC$. They determine a causal domain $D$(while region).}\label{in-out}
\end{figure}

So what does the flux stand for in this picture? It turns out that it is more reasonable if we regard the flux $\int_{C_A} f$ as the logarithm of the rank of $\beta(G,a;\mathcal T)$ \cite{footnote2}. In other words,  we have treated edges on $A$ as inputs and edges on $C_A$ as outputs as shown in FIG.\ref{in-out}, and the flux is given by
\begin{equation}
\int_{C_A} f \equiv \log\ \{rank\ \beta(D_A)\}= \int_{D_A} \rho + \int_{A} f. \label{46}
\end{equation}

Assuming that we have chosen a tensor assignment that makes the $rank\ \beta(D_A)$ maximal. From QMF/QMC conjecture one has:
\begin{eqnarray}
\int_{A} f &=& -\int_{D_{A}} \rho + \int_{C_A} f \nonumber \\
                        &=& -\int_{D_{A}} \rho + \max\ \log\ \{rank\ \beta(D_A)\} \nonumber\\
                &\le& -\int_{D_{A}} \rho + \log\ QMC(D_{A}). \label{a.8}
\end{eqnarray}
The second equality holds when the QMF/QMC theorem is satisfied.

Now let us consider a case where all the cuts ($C_{AB}$, $C_{BC}$,$C_{ABC}$ and $C_{B}$) in (\ref{a.5}) are causal cuts, then it determines a causal diamond $D$. One would define coordinates $(u,v)$ on this network and choose $A$ and $B$ arbitrary as shown in FIG.\ref{in-out}. Then we propose that the flux through this region $D$ defines a volume measure of the network:
\begin{eqnarray}\label{volume}
\int_{D} dV_{network} := \left|\ \int_{D} \rho\ \right|.
\end{eqnarray}

\begin{figure}
\centering
\includegraphics [scale=0.5]{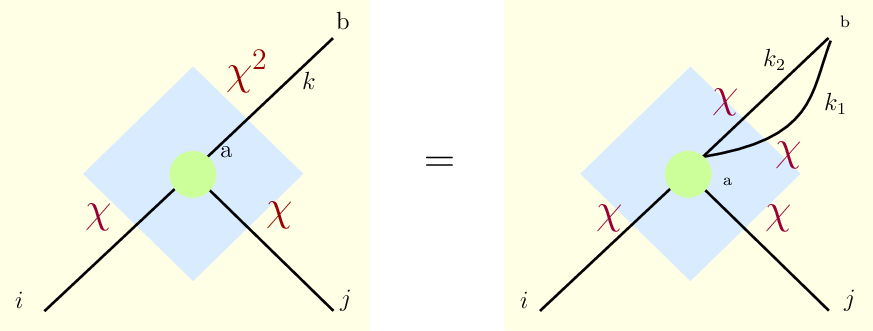}
\caption{A leg $k$ whose Hilbert space dimension is $\chi^2$ can be reshaped into two legs $k_1$ and $k_2$ whose Hilbert space dimensions are $\chi$ respectively.}\label{network}
\end{figure}

Quantum max-flow/min-cut theorem \cite{Cui:2015pla} states that for a tensor network whose Hilbert space dimension of every edge is a power of an integer $\chi$, then QMF=QMC. Hence, in the following considerations, we pay attention to tensor networks where each edge's capacity is a power of integer. In this case, the Hilbert space dimensions of edges associated with a tensor of degree $m$ are given, respectively, by $\chi^{d_1}, \chi^{d_2}, \cdots, \chi^{d_m}$, where $d_1, d_2, \cdots, d_m$ are all nonnegative integers. We can map this graph to a graph of degree $(d_1+d_2+\cdots +d_m)$ such that the Hilbert space dimension of each edge is $\chi$. For instance, consider a simple case $\mathcal T_{ijk}$ where $m=3$, i.e., two input edges and one output edge. The output edge has $\chi^2$ capacity and the rest edges have $\chi$ respectively, see FIG.\ref{network} . Supposing we have a graph that has two parallel edges connecting $a$ and $b$. Clearly there is a one-to-one mapping between left side and right side in FIG.\ref{network}, which preserves the rank and  each tensor $\mathcal T_{ijk}$ in Hilbert space $\bigotimes_{i=1}^{3} \mathbb C^{a_e}$ can be reshaped as $\mathcal T_{ijk_1k_2}$. After decomposition the capacity of each edge of the tensor becomes $\chi$. In a word, any tensor where capacity of each edge is power (maybe different) of integer can be reshaped to a tensor whose capacity of each edge has the same power of the integer.

\subsection{QMF/QMC give the density of compression in MERA} \label{3.2}
Now we are on the point of thinking the potential applications of QMF/QMC to the MERA tensor network. In this case, one thing should be careful.
Generically, given a tensor network, we have the freedom to assign tensors in network. For MERA, however, it is usually homogeneous. We usually put the same isometries and entanglers everywhere. If this is the case, instead of the first version of QMF/QMC, we should adopt the second version of the QMF/QMC. This leads to a problem:  we can not make sure  whether there exits a type of tensor which satisfies QMF$=$QMC. Fortunately, this equality holds asymptotically under specific conditions, as what we will talk in \ref{subsection_5.2}.

Now let us suppose that all edges of the MERA are associated with the same Hilbert space dimension $\chi$, and QMF$=$QMC is satisfied
%\sout{when we consider a large $c$ limit as will be discussed in \ref{subsection_5.2}}.
We now consider an exclusive causal cone of region A as a sub network $D_A$ of the whole network. The edges living on UV cutoff of the network $D_A$ are set to be inputs, while the edges emanating from the exclusive causal cone are regarded as outputs (see FIG.\ref{in-out} for detail).
For MERA network, it is obvious that the edges which are cut by causal cone form a minimum cut set of the network $D_A$ because the number of edges living on a space-like cut is always more than the one living on a causal cut. For simplification, we assume that the number of output edges is $k$ so that the dimension of that Hilbert space(also QMC of this sub network) is equal to $\chi^k$. Then one can simply obtain the quantum max-flow of $D_A$ as $\chi^k$ due to the QMF/QMC theorem.

From section \ref{isometric_tensor}, we know that the entanglement entropy between inputs and outputs for this network reaches its maximum $MEE(D_A)=\log\ QMC(D_A)=\log\ \chi^k=k\log\ \chi$. Clearly, it shows that the entanglement entropy is equivalent to counting the number of edges cut by the causal cut where the weight of each edge is just $\log\ \chi$.
%\sout{From these we can argue that the entanglement entropy can be regarded as a ``flux'' through the causal diamond of an oriented Lorentzian manifold as pointed out in \cite{Czech:2015kbp} and the flux is determined by QMC$(D_A)$. The maximum flux of an edge is equal to $\log\ \chi$. In other words, given an arbitrary edge of a network the corresponding flux has an upper bound $\log\ \chi$.}
That is to say, if a network satisfies the QMF/QMC theorem everywhere then the flux of each edge would achieve its maximum value. Recalling our argument (\ref{a.5}) and (\ref{a.8}) in \Rmnum{2}.A, the equality in (\ref{a.8}) holds because of the QMF/QMC theorem and one can obtain the flux of a causal diamond $D$ by replacing QMF with QMC in(\ref{a.5})
\begin{eqnarray}
\mathcal D &=&  -\left(\int_{D_{AB}} \rho + \int_{D_{BC}} \rho - \int_{D_{ABC}} \rho - \int_{D_B} \rho \right) \nonumber \\
                      &=&  -\rho \left( \int_{D_{AB}} + \int_{D_{BC}} - \int_{D_{ABC}} - \int_{D_B} \right) \nonumber \\
                      &=& -\rho V_D \nonumber \\
                      &=& \log\ \frac{QMC(D_{AB})\cdot QMC(D_{BC})}{QMC(D_{ABC})\cdot QMC(D_{B})}. \label{a.10}
\end{eqnarray}
In the second line $\rho$ is taken out from integral because MERA has same isometries everywhere so it is independent of $(u,v)$.
%\sout{One should be able to notice that in MERA the causal cut is also the min-cut of $D_A$.}
The last line of (\ref{a.10}) can be expressed as $(k_{AB}+k_{BC}-k_{ABC}-k_B)\cdot \log\chi$, where $k_{AB}$ is the number of min-cut edges of $D_{AB}$ and so on.

Obviously, canceling incoming and outgoing flow will yield flux $\mathcal D$. In \cite{Czech:2015kbp} the authors claimed that the number of remaining edges is equal to the number of isometries inside the diamond we consider.
%\sout{Eq. \eqref{a.10} shows that $\mathcal D = \rho V_D$ when the QMF/QMC theorem holds. Recalling the fact that the flux $\mathcal{D}$ is exactly proportional to the number of isometries inside the diamond since the flux of each edge is at most $\log\ \chi$. We thus call $\rho$ the density of isometries.}
This is the case that density $\rho$ is just a constant. And each isometry contains $1/(4G_N)$ bits information. However, $\rho(x)$ is related to location $x$ in tensor network when type of isometries is unconstrained. Different isometry may contains different number of bits and the emerged space-time has totaly different structure. if $\rho$ is constant, from (\ref{a.10}) that the number of isometries is directly proportional to volume. This recovers the argument given in \cite{Czech:2015kbp}. In that article, the density of compression of a compression network is proportional to the number density of isometries for a vacuum MERA.
%\sout{Meanwhile the volume of the causal diamond stands for some conditional information for corresponding regions A, B and C on the boundary as shown in FIG.\ref{network}(b).}
Based on these observations, the authors claimed that $\mathcal{D}(isometries)=I(A,C|B)$, which is a relation between the number of isometries and corresponding volume. This statement is consistent with our argument given above (\ref{a.10}), namely, as the  QMF/QMC theorem holds for  a given network, $\rho V_D$ then can be interpreted as the density of compression. From this point of view, physically, $\rho$ can be viewed as density of isometries or equivalently density of compression.

From FIG.\ref{network}(b), it is obvious that a causal diamond contains numbers of tilted chessboards, each of them corresponds to an isometry. This implies that the volume of every minimum chessboard (or unit chessboard) is same because it contains only one isometry. This property turns out to be the key for the tensor network to have the geometry of $dS_2$. This can be also obtained from (\ref{a.10}) when $D$ is an infinitesimal causal diamond. Indeed, it follows from \cite{Cui:2015pla,Czech:2015kbp} and (\ref{a.10}) that
\begin{align}
\begin{split}
dV_{network} & = \mathcal D = |\rho| V_D = I(A,C|B) \\
             & = 4\log \chi \cdot \frac{dudv}{(v-u)^2}. \label{4.2.2}
\end{split}
\end{align}
The last equality we will obtain in the coming discussion. One can find directly that
\begin{eqnarray}
|\rho| = 2\log\chi.
\end{eqnarray}

\subsection{Holographic entanglement entropy}
In this subsection we return to discuss the holographic entanglement entropy, in the framework of our flow language. Before doing that, we should claim two important properties of this flow which are useful in our following discussion. Consider a tensor network which includes coarse-graining (or isometries), firstly we assume that the Hilbert space dimension of each edge is equal. For such a network, going along renormalization flow each step of coarse-graining will reduce the number of edges. That means in a causal domain $D$ the lower cut number is always greater than the upper cut number. We assume that the flow runs along the RG flow direction. The first property is that the flux of this region $\oint_{\partial D} f$ is always nonnegative, from (\ref{a.5}) it educes
\begin{eqnarray}
\int_D \rho \le 0. \label{3.3.1}
\end{eqnarray}
The second property is deduced from the first property apparently. Suppose we have two regions $A$, $B$ of the boundary, then the flux $\mathcal D_{AB}$ is always greater than the sum of $\mathcal D_A$ and $\mathcal D_B$:
\begin{eqnarray}
\int_{D_{AB}} \rho \le \int_{D_A} \rho + \int_{D_B} \rho. \label{3.3.2}
\end{eqnarray}
Generalization to more than two regions is the same.
%\sout{This inequality implies that there exist some densities of compression of $D_{AB}$ that are not included in $D_A$ and $D_B$.}
This inequality implies that $D_{AB}$ includes some densities of compression which are not included in $D_A$ and $D_B$. Actually these densities provide the conditional mutual information between $A$ and $B$ on boundary as shown in \cite{Czech:2015kbp}.

For a given flow, it is easy to check from the second property that
\begin{eqnarray}
\int_{C_{AB}} f &=& \int_{AB} f - \int_{D_{AB}} \rho \nonumber \\
                      &\le& \int_A f - \int_{D_A} \rho+ \int_B f - \int_{D_B} \rho \nonumber \\
                      &=&\int_{C_A} f + \int_{C_B} f. \label{3.3.4}
\end{eqnarray}

 Now return to the MERA network which exhibits these two properties properly. Then the RT formula can be represented by
\begin{eqnarray}
S(A) = \max\ \int_{C_A} f, \label{3.3.3}
\end{eqnarray}
which is the maximum flux through the causal cut $C_A$. We can simply denote it as $\int_{C_A}f(A)$. Then after using the second property (\ref{3.3.2}), and choosing a flow $f(AB)$ which maximizes the flux through $AB$, we have
\begin{eqnarray}
S(A) + S(B) &\ge& \int_{C_A} f(AB) + \int_{C_B} f(AB) \nonumber \\
            &\ge& \int_{C_{AB}} f(AB) = S(AB).
\end{eqnarray}
This is nothing but the subadditivity of the entanglement entropy.
%\sout{The first property (\ref{3.3.1}) also implies this property, considering the mutual information $I(A:B) := S(A) + S(B) - S(AB) = \int_D \rho \ge 0$, where $D=D_{AB}-D_A-D_B$.}
%More than this, we have known in MERA network this volume describes the conditional mutual information of boundary region.
Similarly, for the case including three regions, we choose a flow $f(A,B,C)$ which maximizes the flux through $A$, $B$, $AB$, $BC$ and $ABC$ simultaneously. Eq.(\ref{3.3.1}) leads to the strong subadditivity of the entanglement entropy
\begin{eqnarray}
I(A:C|B) &=& \int_{C_{AB}} f(A,B,C) + \int_{C_{BC}} f(A,B,C) \nonumber \\
         &-& \int_{C_{B}} f(A,B,C) - \int_{C_{ABC}} f(A,B,C) \nonumber \\
         &=& S(AB)+S(BC)-S(B)-S(ABC) \nonumber \\
         &=& -\int_D \rho \ge 0.\label{57}
\end{eqnarray}

\section{Interpretation}
\begin{figure}
\centering
\includegraphics [scale=0.4]{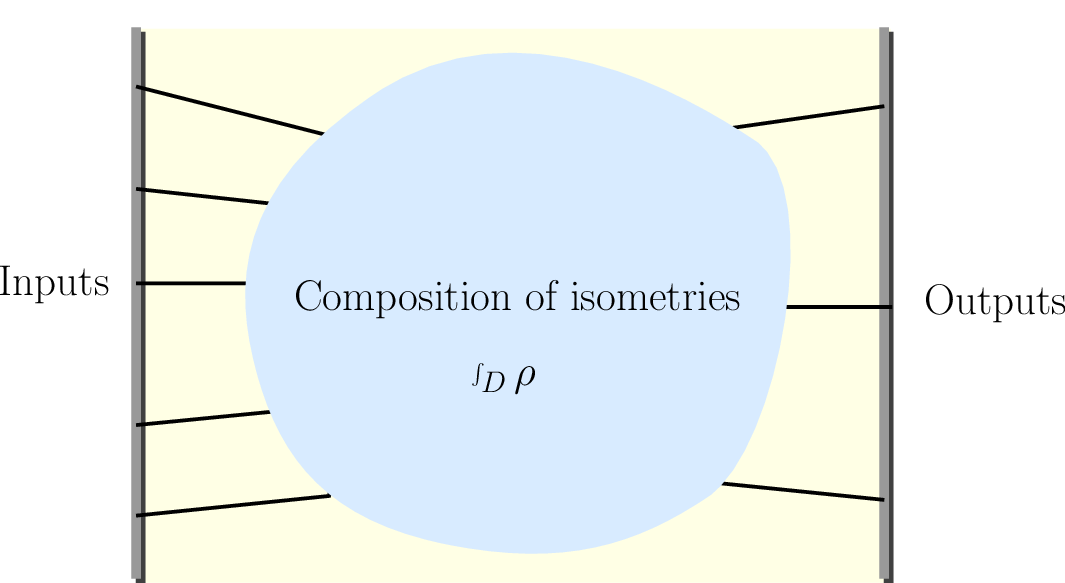}
\caption{For a general tensor network, the emerged manifold is determined by the general density $\rho(x)$.}\label{general_rho}
\end{figure}

In this section we try to give more details about the quantum bit threads model from physical point of view
%\sout{, such as the auxiliary space-time and the role of the central charge $c$ of boundary quantum system.}
. Firstly, we show the meaning of  $\rho(x)$ and some requirements of such local contribution of a certain conditional mutual information. Secondly, we will see because of $\rho =$ constant, the 2D space-time structure constructing by a coarse-graining tensor network is a $dS_2$, which is the same as kinematic space. As mentioned above, to keep the entanglement flow conserved one has to introduce a density $\rho$ of isometry tensor. This can be observed when the QMF/QMC theorem holds. In general, $\rho$ depends on the position of isometry: $\rho(x)$ and related to the structure of tensor network(see FIG.\ref{general_rho}). From (\ref{volume}) the emerged manifold is determined by such density $\rho(x)$. We will see later for the MERA case where $\rho$ is a constant and the emerged space-time is scale invariant(i.e. it is a de Sitter space-time). We can also obtain the relation of Hilbert dimension $\chi$ and central charge $c$ of boundary theory. Finally, we will talk about the role of central charge $c$ in QMF/QMC and space-time.

%\sout{As we mentioned above, To keep the entanglement flow conserved one have to introduce a density $\rho$ of isometry tensor. This can be observed when QMF/QMC theorem holds. In our viewpoint the MERA tensor network is a quantum circuit and the direction of coarse granning can be regarded as the time order of the circuit. Therefore the density $\rho$ describes the degree of compression of isometry along coarse graining. For example, in the cMERA each isometry tensor compress $2\log \chi$ entanglement flow into $\log \chi$ in each layer. In general, $\rho$ is dependent on the position of isometry: $\rho(x)$ and related to the structure of tensor network(see FIG.\ref{general_rho}). From (\ref{volume}) the emerged manifold is determined by such density $\rho(x)$. We will see later for the MERA case $\rho$ is a constant and the emerged spacetime is scale invariant(i.e. it is a de Sitter spacetime).}

\subsection{$\rho$ is the local contribution to conditional mutual information}
In last section we initially set up the quantum bit threads by defining $\rho$ in the tensor network. It can be understood as a source(or sink) that counting how many quantum bits are decompressed(or compressed) in the isometry. In this subsection we try to provide more interpretation of such density in MERA.

Recently a concept called ``entanglement contour'' of quantum systems has been studied in holography \cite{Kudler-Flam:2019oru,Vidal:contour}. In general, the contributions to entanglement entropy come from not only degrees of freedom near the boundary between $A$ and $\bar{A}$, but also degrees of freedom further away from this boundary. The entanglement contour $s_A(x)$ is defined as a local quantity \cite{Kudler-Flam:2019oru,Vidal:contour}:
\begin{equation}
S(A)=\int_{A}s_A(x),
\end{equation}
that aims to quantifying how much degrees of freedom in region $A$ contribute to the total entanglement entropy. In this paper we propose that, similar to the entanglement contour, the density $\rho(x)$ contribute local degrees of freedom between two interval to the total conditional mutual information. As mentioned in section \ref{iii}, the isometries encode density of compression that how much information(q-bits) sharing between two infinitesimal intervals $[u-\Delta u, u]$ and $[v, v+\Delta v]$. In other words, $\rho(x)dV$ encodes the conditional mutual information of local degree of freedom only between local region $[u-\Delta u, u]$ and $[v, v+\Delta v]$. We further find that such density $\rho(x)$  satisfies some requirements similar to those of entanglement contour \cite{Kudler-Flam:2019oru,Vidal:contour}.

The requirements of $\rho$ in the MERA tensor network is following. Here we denote $\rho_{D}(x)$ as the density of conditional mutual information $I(A_1,A_3|A_2)$, where $D$ corresponds to the causal domain in the kinematic space of boundary subregion $A_1\cup A_2\cup A_3\equiv A$. Let us denote $\tilde{\rho}_{D}(x)=-\rho_{D}(x)$. Then the requirements of $\tilde{\rho}_{D}(x)$ are given by:
\begin{itemize}
\item [(1)]
Positivity: $\tilde{\rho}_{D}(x)\ge 0, \forall x\in D.$
\item [(2)]
Normalization: $I(A_1,A_3|A_2)=\int_{D}\tilde{\rho}_{D}(x).$
\item [(3)]
Invariance under symmetry transformations:  Consider four sites $i_1,i_2 \in A_1$ and $j_1,j_2 \in A_3$ and $|j_1-i_1|=|j_2-i_2|$. Let $T$ be a symmetry of reduced density matrix $\varrho_A$, that exchanges two sites $i,j\in A$. After exchanging two points $x_1,x_2\in D$, where $x_1=(i_1,j_1), x_2=(i_2,j_2)$. Then we have $\tilde{\rho}_D(x_1)=\tilde{\rho}_D(x_2)$.
\item [(4)]
Invariance under local unitary transformations: If $\varrho'_A=U_X \varrho_A U_X^{\dagger}$, where $U_X$ is a local unitary transformation supported on $X\subseteq A$, then $\tilde{\rho}_D (K)$ is equal for both $\varrho_A$ and $\varrho'_A$. $K\subseteq D$ is a causal domain corresponding to $a_1\cup a_2\cup a_3$ where $a_1\subseteq A_1$ and $a_3\subseteq A_3$.
\item [(5)]
Upper bound: If we decompose $D$ into causal domain $D_{\Omega}$ and $D_{\Omega}^-$,  and $K$ is contained within $D_{\Omega}$, then
\begin{equation}
~~~~~~~\tilde{\rho}_D (K)\le I(D_{\Omega}), \nonumber
\end{equation}	
where $I(D_{\Omega})$ is the conditional mutual information given by $D_{\Omega}$.
\end{itemize}

In our toy model, requirements (1) and (2) have been discussed in the last section, see (\ref{3.3.1}) and (\ref{57}). The positivity is due to the amount of isometry tensors in the MERA tensor network and $\rho_D (x)$ is regarded as a sink in the network.

Requirement (3) is actually the symmetry of the MERA tensor network. $T$ is a symmetry of $\varrho_A$, i.e., $T\varrho_A T^{\dagger}=\varrho_A$, that exchange two sites. Therefore if we exchange $i_1\leftrightarrow i_2$ and $j_1\leftrightarrow j_2$ at the same time, the system would remain unchanged. This requirement ensures that $\tilde{\rho}_D(x)$ is the same on two points $x_1$ and $x_2$ of $D$. $|j_1-i_1|=|j_2-i_2|$ implies $x_1$ and $x_2$ are two sites in the same layer of MERA tensor network. This reflects the translation symmetry of the quantum system and $\tilde{\rho}_D(x)$ plays an equivalent role in the MERA tensor network.

Requirement (4) refers to subregion K within region $D$. The definition of $\tilde{\rho}_D(K)$ is similar to the one of entanglement contour, which extends the density of single point to the density of a causal domain $K\subseteq D$:
\begin{equation}
\tilde{\rho}_D(K)=\int_{K}\tilde{\rho}_D(x).
\end{equation}
Similar to the entanglement contour \cite{Vidal:contour}, for two disjoint subsets $K_1,K_2\subseteq D$, with $K_1\cap K_2=\emptyset$, the density is additive:
\begin{equation}
\tilde{\rho}_D (K_1\cup K_2)=\tilde{\rho}_D (K_1)+\tilde{\rho}_D (K_2).
\end{equation}
Then if $K_1\subseteq K_2$, the density must be larger on $K_1$ than $K_2$:
\begin{equation}
\tilde{\rho}_D (K_1)\le \tilde{\rho}_D (K_2),  ~~~\text{if}~~~K_1\subseteq K_2
\end{equation}
This is the monotonicity of $\tilde{\rho}_D$. These two properties are observed directly from the kinematic space.

\begin{figure}
\centering
\includegraphics [scale=0.6]{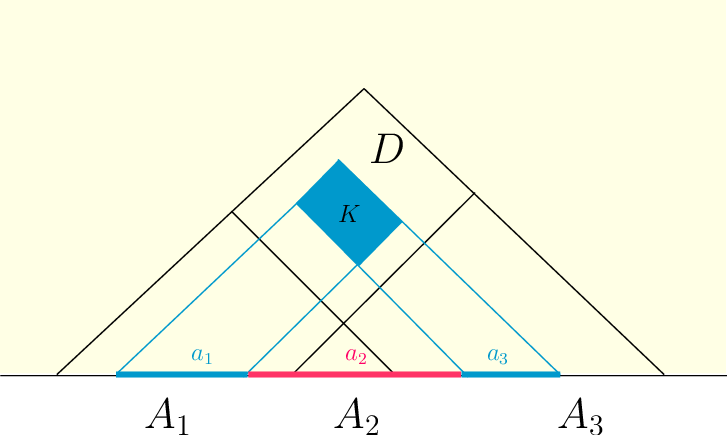}
\caption{}\label{region_K}
\end{figure}

It can be learned from (\ref{46}) that requirement (4) holds in our bit threads language. As we mentioned in section II.B (text below (\ref{QMF=QMC})), after choosing appropriate basis one can let $\beta$ under this basis be matrix $C$. Then reduced matrix $\varrho_A=Tr_{\bar{A}}|\alpha\rangle \langle\alpha|=CC^{\dagger}/Tr(CC^{\dagger})$. The unitary transformation $U_X$ is defined on Hilbert space $V_A\otimes V_{\bar{A}}$ but only nontrivial on $X$. In other words, it does not affect $\bar{X}=A-X\supseteq \bar{A}$. After acting this unitary transformation the reduced matrix is $\varrho'_A=C'C'^{\dagger}/Tr(C'C'^{\dagger}))$. $C'$ is the new matrix under unitary transformation $C'=U_X CU_X^{\dagger}=U_X CU_X^{-1}$, i.e., matrix $C'$ and $C$ have the same rank. Therefore the local unitary transformation does not affect the term $\log \{rank~C\}$ in (\ref{46}). Assuming that the causal domain $K$ in kinematic space is given by $a_1\subseteq A_1$ and $a_3\subseteq A_3$ on the boundary (see Fig.\ref{region_K}). The subregion density on $K$ is given
\begin{equation}
\tilde{\rho}_D(K)=\int_{D_{a_1a_2a_3}+D_{a_2}-D_{a_1a_2}-D_{a_2a_3}}\tilde{\rho}_D(x)
\end{equation}
After using formula (\ref{46}), the second term $\int f$ in the left-hand side of (\ref{46}) will be cancelled so it cannot be affected by the unitary transformation. The remain terms are the rank of matrix $C$, which are unchanged under transformation. In conclusion, the density $\tilde{\rho}_D (K)$ is invariant under unitary transformation.
Requirement (5) is easy to obtain in our bit threads of MERA tensor network. In this case, $\tilde{\rho}$ is a constant and we have
\begin{equation}
\tilde{\rho}_D(K)=\int_{K}\tilde{\rho}_D(x)\le \int_{D_{\Omega}}\tilde{\rho}_D(x)=\int_{D_{\Omega}}\tilde{\rho}_{D_{\Omega}}(x)=I(D_{\Omega}).
\end{equation}

\subsection{The auxiliary space-time}\label{5.A}
The kinematic space of an $AdS_3$ time slice, which is equivalent to an auxiliary $dS_2$ according to the first law of entanglement entropy \cite{deBoer:2015kda, Asplund:2016koz}, can be constructed by the conditional mutual information of a boundary system.

It follows from (\ref{a.10}) that the conditional mutual information can be written as $\log_2\ \chi \cdot (k_{AB}+k_{BC}-k_{ABC}-k_{B})$, where $k_{AB}, k_{BC}\cdots$ are, respectively, the numbers of edges cut by $C_{AB}, C_{BC}$ and so on. For such a coarse-graining tensor network, a region with length $l_{AB}$ satisfies $l_{AB}\cdot e^{-k_{AB}/2} \sim 1$, namely,
\begin{eqnarray}
k_{AB} \simeq 2\log\ l_{AB}.
\end{eqnarray}
Consider the case that regions $A:(u-du,u)$, $B:(u,v)$ and $C:(v,v+dv)$ construct a volume element on $(u,v)$ in the kinematic space, one can obtain
\begin{eqnarray}
dV_{network} &\simeq& 2\log\ \chi \cdot \left[\log\ \frac{(v-u+du)(v-u+dv)}{(v-u+dv+du)(v-u)}\right] \nonumber \\
             &=& 2\log\ \chi \cdot \frac{dudv}{(v-u)^2} + O(dudv),
\end{eqnarray}
which is the conditional mutual information $I(A:C|B) = \partial_u\partial_v S_{ent}(u,v)dudv$. Comparing it with the entanglement entropy of the boundary interval $(u,v)$, i.e, $S_{ent} = (c/3)\log\ (v-u)/\epsilon$, we have
\begin{eqnarray}
\log\ \chi \simeq \frac{c}{6}. \label{4.1.3}
\end{eqnarray}

One thing deserves emphasis. The above discussion is applied in the planar coordinates of $dS_2$. Its topology is a plane $\mathbb R^1 \times \mathbb R^1 $ rather than a cylinder $\mathbb S^1 \times \mathbb R^1$. A plane implies that its cylinder circumference is much larger than the interval, $\varSigma \gg (v-u)$ (actually it is infinite). In other words, if we consider a lattice model on the boundary, the number of sites in $(v-u)$ is much less than those in its complement $(v-u)^c$. If we do not distinguish the direction of geodesics in $AdS_3$ time slice, this kinematic space only covers a half of the full $dS_2$, which is the planar patch $\mathcal O^+$(or $\mathcal O^-$) \cite{SinaiKunkolienkar:2016lgg, vidal_ds/mera}.

However, one should be able to note that the present flow model is a toy model in the sense that the
edges are maximally entangled \cite{footnote3} as we have mentioned in section \ref{isometric_tensor}. In other words, flow in each edge has the same flux and reach their maxima $\log\chi$ simultaneously. This is a strict constraint and only works for some special tensor networks, such as prefect tensor \cite{Pastawski:2015qua} or random tensor \cite{Hayden:2016cfa}.  Recent attempt in constructing a continuous tensor network based on path-integral optimizations \cite{Miyaji:2016mxg, Caputa:2017} may have clues to overcome this difficulty. MERA as an effective simulation for the ground state of real CFT, the entanglement entropy shouldn't be maximal. However, one can control the degree of entanglement in MERA for approximating the real CFT \cite{Bao:2015uaa}. We argue that under this control the corresponding space-time is still dS, but with different density $\rho$, see Appendix \ref{appendix1}.

\subsection{The large c limit}\label{subsection_5.2}
Central charge $c$ is a measure of the number of degrees of freedom. In a strong coupling limit of field theory, when the number of degree of freedom is very large $c\sim N^2\gg 1$ the string interaction becomes weak and we just consider the classical string limit \cite{Maldacena:1997re,tHooft:1973alw}. In other words, in such a limit one can discuss something about the dynamics of quantum CFT by studying a dual semiclassical gravitational physics in space-time.

In (\ref{a.10}) we take out $\rho$ from the integral because we have restricted all the isometries tensor in tensor network in the same type. This corresponds to the second version of QMF/QMC conjecture since we loss the freedom to assign tensors. In this circumstance the QMF/QMC conjecture is not always valid even when each edge has the same capacity, just like the MERA. What we have is a bound for QMF: $QMF(G,a) \le QMC(G,a)$, i.e., flux in each edge can not reach its maximal capacity. This is what a MERA of real CFT needs to has.

Fortunately, recent work from Hastingss \cite{QMF_2} proved that this conjecture is ``asymptotically'' true in the limit as $\chi \rightarrow \infty$. That is to say, the ratio of the QMF to the QMC converges to 1 as $\chi$ tends to infinity. We write $QMF(G, \chi, O)$ to denote the QMF for a given graph $G$ with ordering $O$ and capacity $\chi$ in every edge. Ref. \cite{QMF_2} showed that
\begin{eqnarray}\label{6.4.1}
QMF(G, \chi, O) = QMC(G, \chi)\cdot \left(1-O(1)\right).
\end{eqnarray}
The higher-order term $O(1)$ we consider as an asymptotic function of $\chi$, which may also depend on $G$ and $O$. From (\ref{4.1.3}) this implies the QMF/QMC conjecture is asymptotically true in a large central charge limit. The entanglement entropy $S(A) = \int_{C_A} f$ is asymptotically equal to $\log\ QMC$ and we can just count the number of cut legs. We therefor have a dual classical, at least semiclassical gravitational theory in the auxiliary $dS_2$ space-time. Things can be simplified in such large $c$ limit and  computations of entanglement entropy can be made by a holographic map to a volume in auxiliary space.

We have shown that max-flow/min-cut in classical network is always valid. This, however, fails for a tensor network: the quantum version of max-flow/min-cut conjecture does not hold in general except for large $\chi$. So this result (\ref{6.4.1}) indicates that a quantum phenomenon (QMF $\neq$ QMC) disappears in the large system limit. This is our foothold that the classical space-time can be emerged from the tensor network model. The network becomes ``classical'' and we can define the flow on it, which construct the gravitational theory. This agrees with the argument that in the large $c$ limit, the quantum fluctuations of emergent space-time can be effectively suppressed. This can be noticed from the relation $c\sim\frac{1}{G_{N}}$. Remember that we consider the large $c$ limit because we adopted the second version of QMF/QMC whose tensor was restricted. This will result a nontrivial complexity of the MERA circuit. The duality between tensor network and gravity in our model is supported from the complexity's point of view, which explains that the complexity corresponds to the gravitational action in gravity side \cite{Chen:2018ody}. The quantum corrections may be considered in our toy model of tensor network. These corrections correspond to corrections in the holographic entanglement entropy. We will give a brief discussion in the next section.

Recently the quantum Hamiltonian complexity has been made rich connections with physical system. Physicists concern about some properties of local Hamiltonian in condensed system, such as the ground state properties or entanglement properties. In \cite{Cui:2015pla} it was shown that the quantum max-flow is related to the so-called quantum satisfiability problem, $QSAT$, which is defined as the quantum version of $k-SAT$ in \cite{QSAT}. It was found that in some specific cases the problem of QSAT and QMF are equivalent. So they give a conjecture similar to one in QSAT that the QMF/QMC conjecture holds when the Hilbert space dimensions of edges become very huge.
This conjecture was proved in \cite{QMF_2}. On the other hand, machine learning is also related to tensor network through renormalization in condensed matter physics \cite{dl&rg} and possibly have significant holographic meanings \cite{holo_dl}\cite{holo_ml}. It was shown in \cite{qmfmc_dl} that the quantum max-flow provides a non-trivial measure of the ability of tensor network to model correlations in a so-called deep convolutional network. All of these imply a possible meaning of our understanding about emergent gravity.

\section{Conclusions and discussions}
By making use of the QMF/QMC theorem developed recently in tensor network, we have proposed a tensor network/flow correspondence, which is a quantum generalization of the flow description of the RT formula in \cite{bit-threads}. Based on information-theoretic considerations, we suggest that for MERA we need to introduce a new variable $\rho$, which is interpreted as the density of the tensor networks. Physically, this term is viewed as the source (or sink) of the tensor networks thus plays a significant role in the flow description of the RT formula. $\rho(x=(u,v))$ is a local contribution to a certain conditional mutual information between local degrees of freedom only in sites $u$ and $v$. Its value is not only closely related to the geometric structure of the tensor network (equivalently, the metric of emergent space-times from the emergent point of view), but also the density of compression or decompression of quantum bits through reducing or expanding the dimension of Hilbert space, which implies a naive picture that the evolution of our universe can be regarded as a huge and complex quantum circuits. Density $\rho$ is the main role for the space-time's construction in the large $c$ limit. The MERA circuit is a program which entangle quantum bits continuously, and the complexity $\mathcal{C}$ can be given by counting the number of entangled pairs in isometry in such limit,
\begin{equation}
\mathcal{C} \sim \int \sqrt{-g}\rho.
\end{equation}
We can relate this complexity to the gravitational action of the de Sitter space and $\rho$ plays the role of gravitational constant \cite{Chen:2018ody}.

Our proposal of quantum bit threads provides a new perspective on holographic tensor network and is different from the one of Headrick-Freedman. The bit threads of Headrick-Freedman are constructed in an AdS time slice. However, we propose the quantum bit threads in the kinematic space instead of its original AdS time slice. We would like to emphasize that the classical limit of our quantum version of bit threads is not the classical network of Headrick-Freedman's because of the differennce between quantum network and classical network (one can found more details of definition in \cite{Cui:2015pla}). %The QMF=QMC case is the classical limit of quantum network but do not consistent with the classical network.

One can further consider the $1/N$ quantum corrections to the quantum bit threads. The quantum corrections of holographic entanglement entropy have been studied \cite{Faulkner:2013ana}. The QMF is asymptotically equal to QMC in the large $c$(or small $1/N^2$) limit, as shown in (\ref{6.4.1}). QMF is calculated by the input-to-output map $\beta$, or equivalently the matrix $C$. The entanglement entropy can be also obtained by matrix $C$ as shown in (\ref{EE}). Hence the quantum corrections of quantum bit threads can be also studied in the way similar to the quantum corrections of HEE. We hope more details of this quantum effect of bit threads can be studied in the future.

\appendix
\section{MERA of the ground state of real CFT}\label{appendix1}
If the tensor network is MERA, we know the entanglement entropy of the state with $l$ sites has upper bound
\begin{equation}
S_{\text{MERA}}(l)\le4f(k)\log_{k}l\cdot\log\chi,
\end{equation}
where the interval is larger than the lattice spacing i.e. $l\gg1$. $\chi$ is the the Hilbert dimension of each edge, $k$ is the number of sites in a block to be coarse-grained and $f(k)$ is a function of $k$ with $f(k)\le k-1$. $\log\chi$ is the maximum entanglement entropy of a single edge when we trace out the rest of the MERA. It's instructive to introduce a parameter $\eta\in[0,1]$ to describes the degree of entanglement \cite{Bao:2015uaa}. In other words, the average entanglement entropy per edge in MERA is $\eta\log\chi$. Then one can write the entanglement entropy as
\begin{equation}\label{MERAentropy}
S_{\text{MERA}}(l)=4f(k)\log_{k}l\cdot\eta\log\chi.
\end{equation}
Recall the entanglement entropy of CFT $S(l)=(c/3)\log l$, then the MERA entropy (\ref{MERAentropy}) gives a central charge
\begin{equation}\label{MERAcentralcharge}
c=\frac{3L}{2G_{N}}=12\eta f(k)\frac{\log\chi}{\log k}.
\end{equation}
Because each edge has the same average entanglement entropy $\eta\log\chi$ we still can count the number cut by causal cut for obtaining the entanglement of region $l_0$. The auxiliary space-time given by MERA is still a de Sitter, but with relation between $\chi$ and central charge $c$ (\ref{MERAcentralcharge}) different with (\ref{4.1.3}).

\section{Comparing with entanglement density}
Another thing we should point out is that in \cite{Nozaki:2013wia} the authors define the entanglement density in $(1+1)$-dimensional CFT. The entanglement density $n(l, \xi, t)$ is defined by counting the number of entanglement pair between the two points $x=\xi-l/2$ and $x=\xi+l/2$. After comparing with the result of entanglement entropy in CFT$_2$ they find
\begin{equation}
n_{CFT}(l,\xi,t)=\frac{c}{6l^2},
\end{equation}
where $c$ is the central charge and $l$ is the range of subsystem we consider. The horizontal ladders are the disentanglers which are unitary transformations between two spins. Each disentangler carries the entanglement entropy $\log 2$ between these two spins. The number of disentanglers in each bond $N(l,\xi,t)$ can be roughly given by \cite{Nozaki:2013wia}
\begin{equation}
N(l,\xi,t) \simeq n_{CFT}(l,\xi,t)\cdot l^2 = \frac{c}{6},
\end{equation}
which can be understood as the density of disnentanglers.

Although this expression of density of disentanglers looks very similar to our density of compression, $\rho\propto N(l,\xi,t)=c/6$. We claim they have different meaning. First, in \cite{Nozaki:2013wia} the MERA is defined in a time slice of AdS space. But in argument of \cite{Czech:2015kbp}, which is the premise of our paper, the MERA lives on the kinematic space rather than the time slice of AdS. The kinematic space is a dS$_2$ which has a natural causal relation between tensors. Second, $N(l,\xi,t)$ is the density of disentanglers in each bond but $\rho$ is the density of compression in each isometry. Third, $N(l,\xi,t)$ is the number of disentanglers in each bond, which describe how many entanglement pair in such a bond. So for obtaining the entanglement entropy of a subsystem one can roughly count the number of intersecting bonds and multiply the density $N(l,\xi,t)$. On the other hand, in our paper we defined the density $\rho$ from an information point of view. The volume of kinematic space is the conditional mutual information which shares the information between two regions. The $\rho$ describes how many information will be shared, or be compressed along the quantum circuit, which have different meaning with the density of disentanglers. For obtaining the conditional mutual information we can count the number of isometry and multiply the density $\rho$.

\section*{\bf Acknowledgements}
We would like to thank Tadashi Takayanagi for reading and commenting
on the paper. We are also grateful to Wen-Cong Gan and Bo Xiong for useful discussions.This work was supported in part by the National Natural Science Foundation of China under Grant Nos. 11975116, 11665016 and 11563006, and Jiangxi Science Foundation for Distinguished Young Scientists under Grant No. 20192BCB23007.


\begin{thebibliography}{42}
\bibitem{footnote1}
  Properly speaking, it refers to a special tensor network---the multi-scale entanglement renormalization ansatz (MERA) which was introduced in \cite{vidal}.
\bibitem{footnote2}
  Detailed explanation will be made in Section \ref{subsection_5.2}.
\bibitem{footnote3}
  We would like to thank T. Takayanagi for bringing this point to our notice.
\end{thebibliography}
\end{document}